# Quantitative assessment of the microstructural factors controlling the fatigue crack initiation mechanisms in AZ31 Mg alloy

Abbas Jamali[a,b], Anxin Ma[a], Javier LLorca[a,b],*

[a] *IMDEA Materials Institute, 28906 Getafe, Madrid, Spain*
[b] *Department of Materials Science, Polytechnic University of Madrid/Universidad Politécnica de Madrid, E. T. S. de Ingenieros de Caminos, 28040 Madrid, Spain*

*Corresponding author: javier.llorca@upm.es, javier.llorca@imdea.org

## Abstract

The deformation and fatigue crack nucleation mechanisms were studied by means of slip trace analysis and secondary electron microscopy in a textured AZ31B-O Mg alloy subjected to fully-reversed cyclic deformation at two different cyclic strain semi-amplitudes. Samples were deformed in two orientations leading to symmetric and non-symmetric cyclic stress-strain curves due to the activation of different deformation mechanisms. They were ascertained in longitudinal sections of the specimens, which included a large number of grains (from 1500 to 4500 for each specimen), to obtain statistically significant results. If the dominant deformation mechanisms were basal slip and tensile twinning/detwinning, the most damaging fatigue cracks were nucleated along twins in large grains, together with cracks parallel to basal slip bands associated with the localization of deformation in clusters of small grains suitably oriented for basal slip. If the main deformation mechanisms were tensile twinning/detwinning and pyramidal slip, the longest fatigue cracks were nucleated along pyramidal slip bands in large grains. Grain boundary cracks around small grains were found in all cases, but they were not critical from the viewpoint of fatigue failure. This information is relevant to assess the effect of the microstructural features on the fatigue life of Mg alloys and as input to simulate the fatigue behavior of Mg alloys using fatigue indicator parameters.

*Keywords*: Fatigue; Deformation mechanisms; Crack initiation; Magnesium alloys; Slip trace analysis.

# 1. Introduction

Fatigue failure due to the nucleation and growth of cracks is one of the most important sources of damage in materials used in structural applications [1]. Fatigue cracks appear due to the progressive localization of deformation at particular zones in the microstructure, which leads to the development of irreversible strains (even when the far-field applied stresses or strains are in the elastic regime) and eventually to the nucleation of cracks. Stress concentrations associated with manufacturing defects (such as voids in metallic alloys or delaminations in composites) or with the presence of large intermetallic particles are often identified as crack initiation locations. However, these defects can be eradicated with improved manufacturing processes. The ultimate factors that trigger the formation of fatigue cracks (and, ultimately, determine the fatigue limit) can be traced to the inherent heterogeneities associated with the microstructural features. For instance, cyclic deformation of alloys containing precipitates that can be sheared by dislocations leads to progressive reduction in the precipitate size until they become unstable. Subsequent cyclic deformation leads to the localization of the deformation in persistent slip bands and to the appearance of stress concentrations at the intersection of the slip bands with the surface [2]. In other metallic alloys, crack nucleation is controlled by the presence of very large grains, which are softer and facilitate the formation of slip bands [3]. In the case of Ni-based superalloys, crack nucleation in defect-free microstructures is associated with the localization of slip in bands parallel to annealing twins within large grains due to the mismatch in elastic constants between the parent grain and the twin [4].

Most investigations to ascertain the microstructural factors that control the nucleation of fatigue cracks have been carried out in cubic materials, which are plastically more isotropic than hexagonal ones. Research on hexagonal alloys, such as Ti, is more limited and shows that grain-to-grain interactions may lead to considerable stress concentrations due to the plastic anisotropy of these materials and control crack nucleation [5,6]. These phenomena may be particularly relevant in the case of Mg alloys, which present the highest plastic anisotropy due to the significant differences in the critical resolved shear stress between basal and non-basal slip systems. Moreover, twinning is easily activated in suitably oriented grains in Mg alloy polycrystals and is followed by detwinning during reversed deformation, leading to continuous twinning-detwinning during cyclic loading [7–21].

The fatigue behavior of Mg alloys has been analyzed in different investigations and particular attention has been paid to the dominant deformation mechanisms as a function of orientation. Wrought Mg alloys usually present a strong basal texture along the rolling/extrusion direction, which facilitates the activation of twinning during compression in this orientation, while activation of basal slip is hindered because basal planes are parallel to the extrusion axis. As a result, a strong tension-compression asymmetry in the cyclic stress-strain curves appears due to the significant differences in the critical resolved shear stresses necessary to activate pyramidal slip (dominant during the tensile part of the cycle) and tensile twinning (which controls plastic strain in compression) [7,14,16,19,22–30]. This asymmetry increases with the applied strain amplitude due to the substantial hardening associated with pyramidal slip [16,28,29,31–35]. It decreases if the textured Mg alloys are deformed under fully reversed cyclic deformation in an orientation between 20º-70º between rolling and normal directions [16,28], reaching a mean stress close to 0 at 45º [16]. In this orientation, plastic deformation takes place by basal slip and tensile twinning/detwinning during the tensile and compressive parts of the fatigue cycle. Similarly, the tension-compression asymmetry disappears in Mg alloys with weak texture [12,13,36].

Obviously, the activation of different dominant deformation mechanisms as a function of texture, orientation, and strain amplitude has an important effect on the fatigue crack initiation mechanisms [13,17,32,34,37–40] and on the fatigue life of Mg alloys [16,28]. The experimental analyses reported in the literature indicate that fatigue cracks are nucleated at basal slip bands at low cyclic strain semi-amplitudes ($\Delta\varepsilon/2 \leq 0.5\%$) in textured Mg alloys deformed along the extrusion direction [32,34,37]. On the contrary, fatigue crack initiation is reported to switch to grain boundaries [13,17,34,37] and twins in these circumstances when the cyclic strain semi-amplitude increases. Other investigations have reported that fatigue cracks are nucleated around intermetallic particles [40]. However, very few studies provide statistically reliable quantitative information about the prevalence of each mechanism and the influence of the microstructural factors (grain size and orientation, grain boundary misorientation, etc.) on fatigue crack nucleation. For instance, Deng et al. [17] analyzed over 200 fatigue cracks nucleated in a weakly textured Mg – 3 wt. % Y alloy deformed at a cyclic strain semi-amplitude of 1% and found that 80% were nucleated at grain boundaries while 20% were parallel to slip bands. More recently, Shi et al. [13] studied a limited number of fatigue cracks along rolling direction (RD), normal direction (ND), and 45º RD-ND (29 cracks along RD, 27 cracks along ND, and 43 cracks along 45º RD-ND) in the RD-ND plane of the textured

AZ31 Mg alloy deformed at a cyclic strain semi-amplitude of 1%. They reported that intergranular cracking was the primary crack initiation mechanism along ND and 45º RD-ND, while transgranular cracking was dominant along RD. However, definite conclusions could not be obtained because of the limited data set and, in addition, the effect of microstructural features (grain size, grain boundary orientation) was not studied. In addition, we previously, analyzed the fatigue crack initiation mechanisms in 500 cracks in a rolled AZ31B Mg alloy deformed at a large cyclic strain semi-amplitude (2%) along the RD and reported the influence of grain size and grain boundary misorientation on the probability of intergranular and transgranular cracking [7]. However, there is no statistically reliable information in the literature to assess the influence of different dominant deformation mechanisms (either basal slip, twinning/detwinning of pyramidal slip) on fatigue crack initiation of Mg alloys. In this study, two different strain amplitudes and orientations were selected to explore the effect of the deformation mechanisms on fatigue crack initiation in AZ31 Mg alloy. Therefore, the information obtained on AZ31 Mg alloy on different orientations and strain amplitudes can be extended to other Mg alloys in so far, the dominant deformation mechanisms are equivalent. This information is essential to design Mg alloys with optimum fatigue resistance and also as input to predict the fatigue life of Mg alloys from fatigue indicator parameters [9,36,41].

## 2. Material and experimental techniques

A rolled AZ31B-O Mg alloy slab of 65 mm in thickness perpendicular to the rolling direction was purchased from Magnesium Elektron. The nominal chemical composition (in wt. %) provided by the supplier is 2.89% Al, 1.05% Zn, 0.42% Mn, and balance Mg. Electrical discharge machining was used to prepare specimens with a rectangular cross-section for the fatigue tests. The dimensions of the specimens are depicted in Fig. 1. They were oriented in the rolling direction (RD) and at 45° between rolling and normal directions (45° RD-ND). Specimens with rectangular cross-section were selected in this investigation (instead of cylindrical ones that are standard to measure the fatigue life) to be able to assess accurately the deformation and fracture mechanisms on the flat lateral surfaces during interrupted fatigue tests. Thus, the fatigue lives reported in this paper only provide qualitative information about the effect of orientation and cyclic strain amplitude on this magnitude.

The specimen surfaces were manually ground using abrasive SiC papers of 1200, 2000, and 4000 grit sizes. Afterward, they were polished in four steps with 6 μm, 3 μm, 1 μm, and 0.25 μm diamond paste and cleaned by immersion in an ultrasound bath of pure ethanol for 3 minutes. The specimens were etched using a solution of 0.4 g of picric acid, 2.5 ml of acetic acid, 2 ml of distilled H$_2$O, and 4 ml of ethanol during 10 seconds to reveal the grain boundaries. The grooves formed at the GB were shallow (< 250 nm) and facilitated to discern whether cracks were nucleated at the GB or nearby from the SEM micrographs in thousands of grains. It should be noted that the grooves at the GBs will significantly affect the fatigue crack nucleation mechanisms under high cycle fatigue. However, fatigue crack nucleation in Mg under low cycle fatigue is controlled by the stress concentrations associated with the plastic deformation mechanisms and the effect of the grooves at the GB can be neglected [32,34,37]. Afterwards, the surfaces were again etched with Nital 15% (15% nitric acid and 85% ethanol) during 2 seconds to get a mirror-like surface. Thus, the microstructural details, as well as the interaction between deformation mechanisms (slip bands, twins) with the microstructural features, could be accurately ascertained by scanning electron microscopy (SEM), and simultaneously grain orientation and twins could be revealed by electron back-scattered diffraction (EBSD).

The microstructure of the samples was analyzed with a FEI Helios Nanolab 600i scanning electron microscope equipped with a Nordlys EBSD detector (Oxford Instruments). EBSD measurements were performed with the step size of one-twelfth of the average grain size, an acceleration voltage of 20 kV, and an acceleration current of 2.7 nA. The EBSD maps (including between 1500 to 4500 grains for each sample) were collected from the area at the center of the samples. The MTEX software was used to extract grain size distribution and crystal orientation from the raw EBSD data [42]. The texture of the samples was analyzed by X-ray diffraction. Measurements were performed using Cu K$\alpha$ radiation at 50 kV with the sample tilt angle in the range of 0° to 90° to obtain the orientation of the (0002) basal plane on the transverse direction (TD)/45º RD-ND and RD/TD planes of the material.

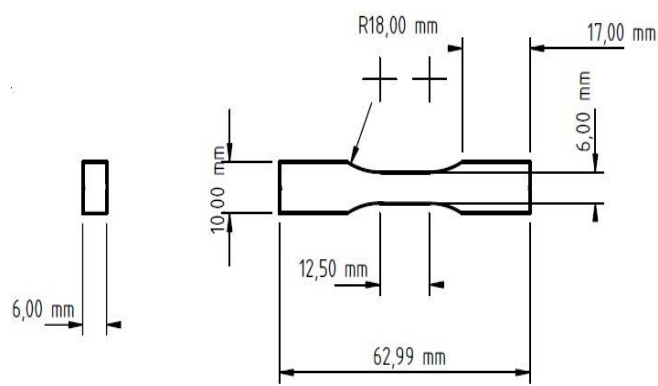

**Fig. 1.** Dimensions of the specimens for the fatigue tests.

Fully-reversed cyclic deformation tests were performed under strain control at three different cyclic strain semi-amplitudes of Δε/2 = 2.0%, 0.8%, and 0.4% in RD and 45° RD-ND in a servo-hydraulic Instron 8802 mechanical testing machine. The strain was controlled with an extensometer attached to the sample and all the samples were initially deformed in compression at the test frequency of 0.5 Hz. Load and deformation were recorded using a computer-controlled data acquisition system. Two tests were performed until failure for each strain amplitude and orientation.

Two different cyclic strain semi-amplitudes of Δε/2 = 2.0% and 0.4% in RD and 45° RD-ND were selected for interrupted tests to have all the dominant deformation mechanisms (basal slip, twinning/detwinning, pyramidal slip) in fatigue of magnesium alloys. These tests were stopped before failure at the number of cycles indicated in Table 1. They correspond to approximately 1/5 of the fatigue life in the specimens oriented at 45º RD-ND and 1/3 of the fatigue life in those deformed along the RD. The fatigue tests were always stopped after the maximum compressive strain was attained. The specimens were taken from the machine and the gage length surface was observed by SEM using EBSD and secondary electron detectors to analyze the deformation mechanisms and the fatigue crack initiation sites.

Table Error! No text of specified style in document.**1.** Number of fatigue cycles in the interrupted tests.

| Orientation | Δε/2 = 0.4% | Δε/2 = 2.0% |
|---|---|---|
| 45° RD-ND | 1200 cycles | 50 cycles |
| RD | 500 cycles | 50 cycles |

## 3. Results and discussion

### 3.1 Microstructure

The EBSD maps of the rolled AZ31 Mg alloy in TD/45º RD-ND and RD/TD planes are shown in Figs. 2a and c, respectively, together with the corresponding inverse pole figures of the (0002) orientations obtained by X-ray diffraction, which are plotted in Figs. 2b and 2d. The c axis of most grains forms an angle of ≈45º with the TD/45º RD-ND plane (Fig. 2b) and, thus, they are suitably oriented to accommodate plastic deformation by basal slip and twinning when deformed along the 45º RD-ND in this plane. On the contrary, the material has a strong basal texture in the RD-TD plane with the c axis of most grains perpendicular to the RD-TD plane (Fig. 2d). Thus, the basal slip will be hindered by the low Schmid factor of all basal slip systems due to the orientation of the basal planes and tensile twinning can only take place in compression in most of the grains during deformation along the RD. The grain boundary misorientation distributions for both planes (obtained from the EBSD maps) are plotted in Fig. 2e and they are practically superposed with a maximum of around 30º. There is a very small peak in both cases at around 86º, which indicates that there were very few twins in the as-processed alloy. The cumulative grain size distributions in both planes (also obtained from the EBSD maps) are plotted in Fig. 2f. The average grain sizes are 7.2 ± 4.4 µm (TD/45º RD-ND) and 9.1 ± 5.2 µm (RD-TD) and this difference can be attributed to the rolling process. Overall, more than 80% of the grains were < 10 µm in both orientations but large grains (reaching up to 50-60 µm) were found in both cases. The fraction of large grains was higher in the RD/TD plane.

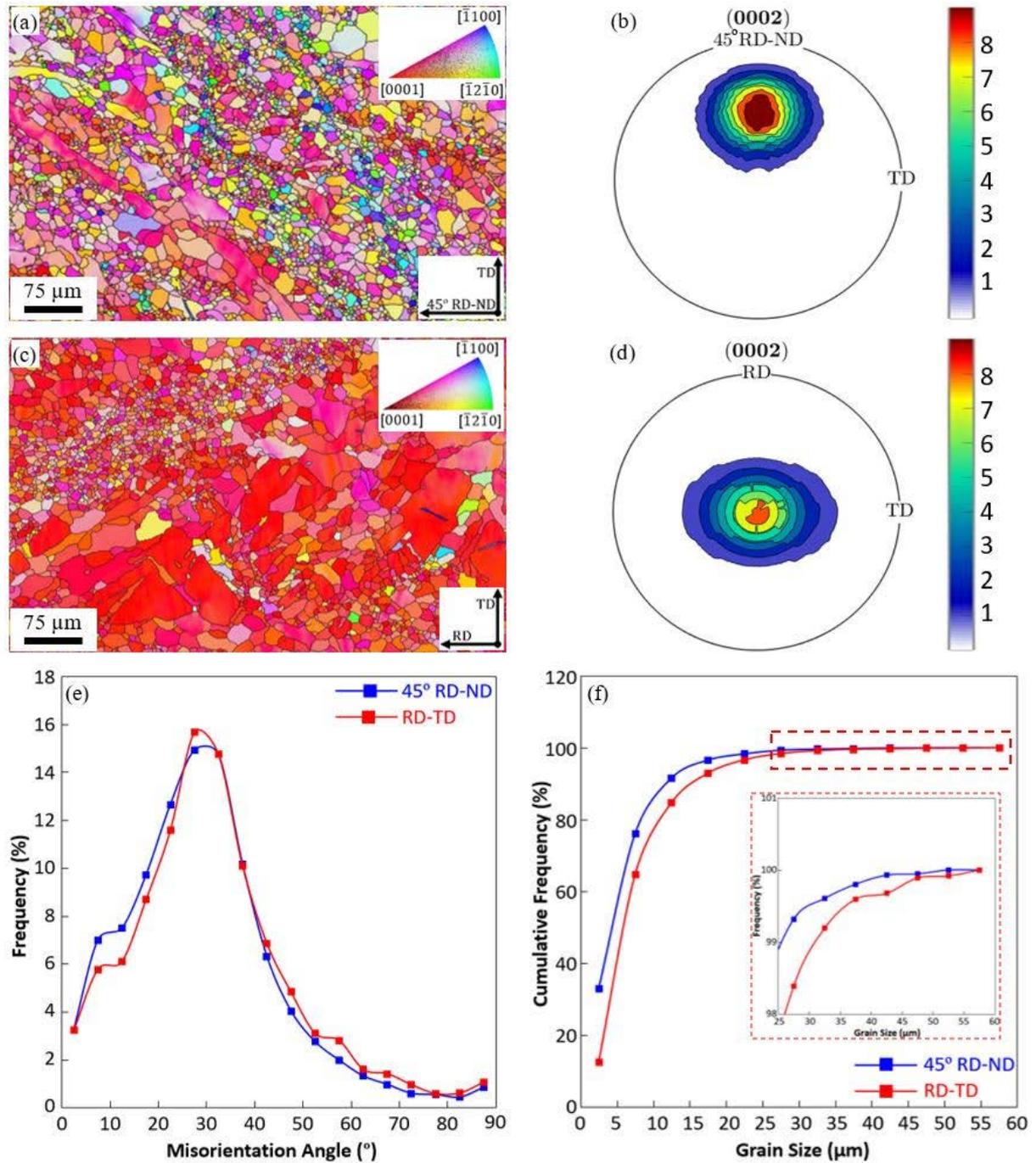

**Fig. 2.** (a) EBSD image and (b) XRD pole figure of the (0002) basal plane showing the orientation of the grains corresponding to the 45º ND-RD/TD plane. (c) and (d) *Idem* corresponding to the RD/TD plane. (e) Grain boundary misorientation distribution for both planes. (f) Grain size cumulative distribution for both planes. Numbers in (b) and (d) stand for multiples of random distribution. The reference directions of the IPF maps are perpendicular to the paper.

3.2 Fatigue tests

The cyclic stress-strain curves of the specimens deformed along 45º RD-ND during the first cycle and at half of the fatigue life are plotted in Figs. 3a and c, respectively. Those for the specimens deformed along the RD are plotted in Figs. 3b and 3d. The hysteresis loops are symmetric from the first fatigue cycle in the specimens deformed along 45º RD-ND and this behavior indicates that similar deformation mechanisms are active in the tensile and compressive parts of the fatigue cycle. Taking into account the orientation of most of the grains (Fig. 2b), basal slip is expected to be the dominant deformation mechanism together with tensile twinning [11–13,16]. In addition, grains twinned during the compressive part of the cycle will detwin in the tensile part and viceversa. On the contrary, the hysteresis loops of the specimen deformed along the RD are not symmetric. Plastic deformation during the compression part of the first cycle occurs at constant stress (Fig. 3b) and this behavior is in agreement with massive twining, which is supported by the strong basal texture (Fig. 2d). Detwinning is expected to be active during unloading, particularly when the far-field stresses become tensile until it is fully completed. Tensile twinning cannot accommodate further plastic deformation in tension and most grains are not suitably oriented for basal slip. Thus, pyramidal slip has to be activated, leading to a strong strain hardening (Fig. 3b) [22,25,41,43,44]. As a result, the maximum stress in tension in each fatigue cycle is higher than the maximum stress in compression and the asymmetry increases with $\Delta\varepsilon/2$ because of the strain hardening associated with pyramidal slip [16,28,29,31–35]. Taking into account the corresponding shapes of the hysteresis loops at half of the fatigue life for each orientation (Figs. 3c and d), the plastic deformation mechanisms seem to be constant during the whole fatigue life for each orientation and cyclic strain semi-amplitude.

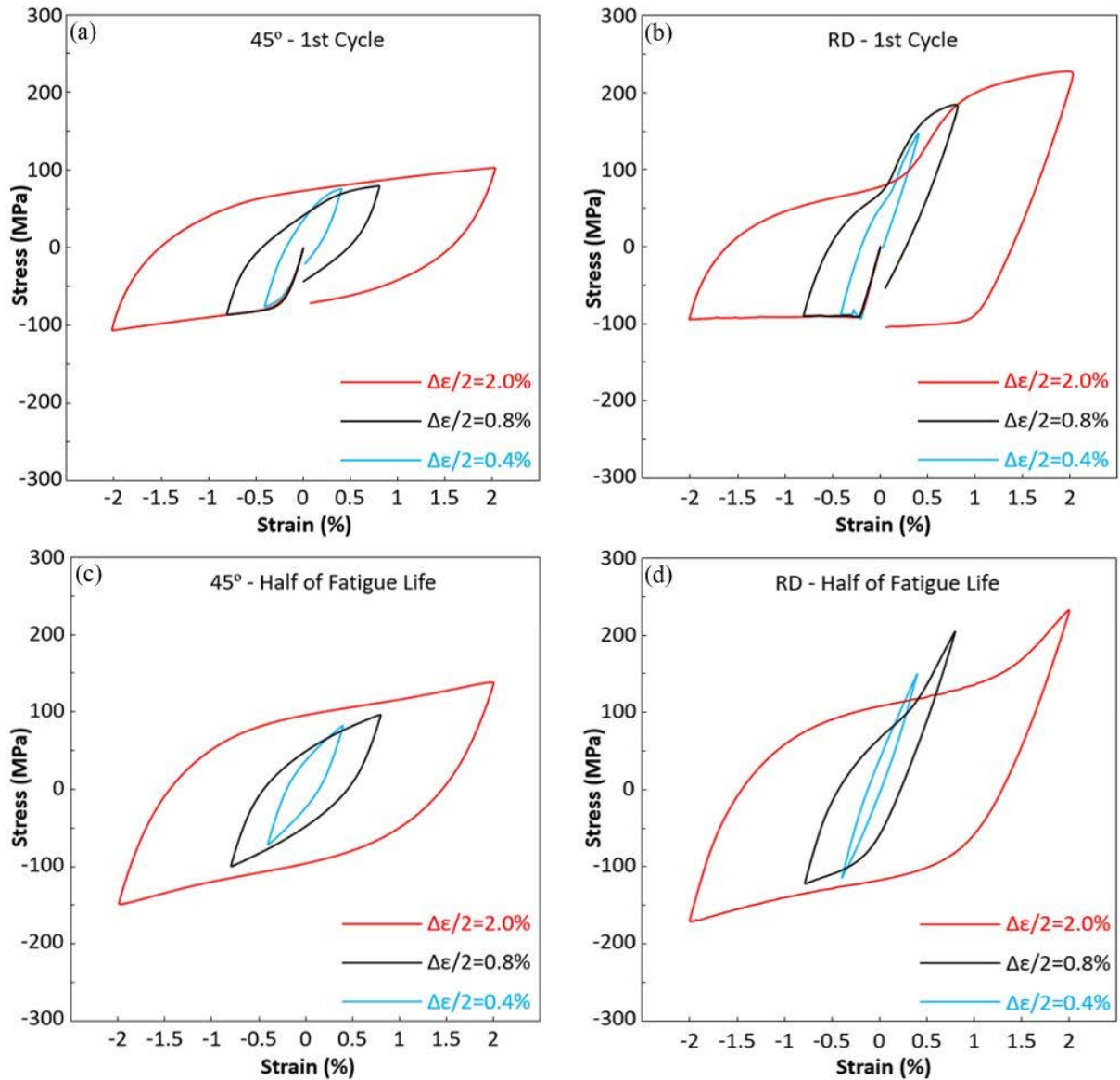

**Fig. 3.** Cyclic stress-strain curves as a function of the applied strain semi-amplitude Δε/2 for different orientations. (a) First cycle, 45º RD-ND. (b) First cycle, RD. (c) Half of the fatigue life, 45º RD-ND. (b) Half of the fatigue life, RD.

The maximum and minimum stresses are plotted in Figs. 4a and 4b for the 45º RD-ND and RD orientations, respectively. The magnitude of the maximum and minimum stress is the same in the case of 45º RD-ND, while the maximum tensile stresses are higher than the compressive ones along RD, and the differences increase with Δε/2. In general, the maximum and minimum stresses were constant through the fatigue life at Δε/2 = 0.4% and a slight cyclic hardening was observed at higher Δε/2 in all cases. The largest cyclic hardening was observed in the compressive stress of the specimens deformed along the RD (Fig. 4b) and could be attributed to the latent hardening associated with the interaction of the twin boundaries with the pyramidal dislocations accumulated during the tensile part of the cycle. Finally, the fatigue life is plotted vs. Δε/2 for both orientations in Fig. 4c. The fatigue life at 45º RD-ND is much longer than that along the RD. The origin of this difference might be associated with the higher strain energy dissipation (because of the larger area of the hysteresis loops in the cyclic stress-strain curve along RD) and the higher magnitude of the maximum tensile stress in each fatigue cycle along the RD due to the activation of pyramidal slip [13]. Moreover, larger number of twins appear during deformation along the RD, facilitating the nucleation of long fatigue cracks [45].

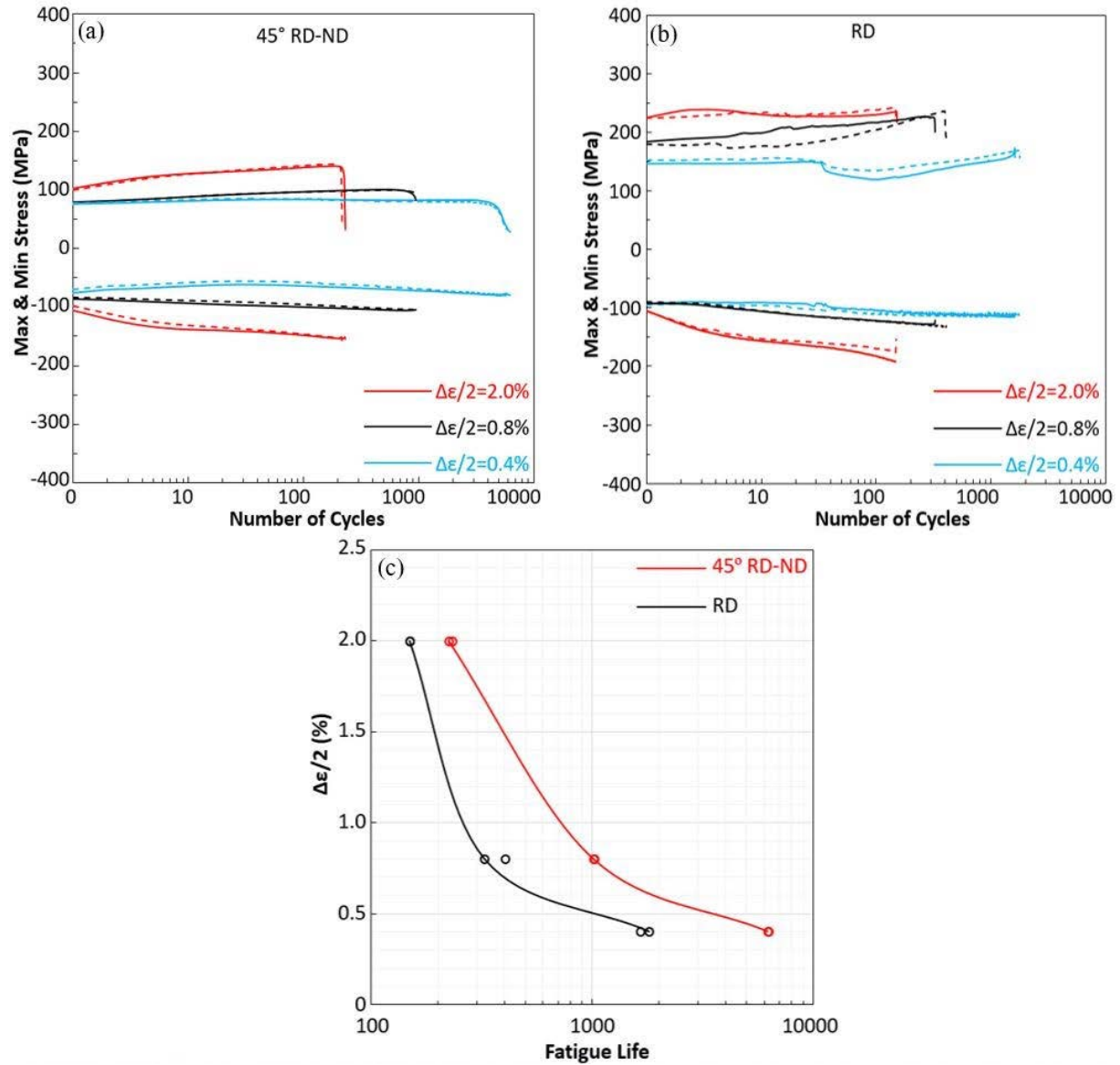

**Fig. 4.** (a) Evolution of the maximum and minimum stress in each fatigue cycle in specimens deformed at different Δε/2 along 45º RD-ND. (b) Idem along the RD. (c) Fatigue life as a function of Δε/2 for both orientations.

3.3 Cyclic deformation mechanisms

Evidence of the dominant deformation mechanisms as a function of the loading orientation and strain amplitude was obtained from the analysis of the surface of the specimens after 1/5 of the fatigue life (45º RD-ND) or 1/3 of the fatigue life (RD) (Table 2). EBSD maps showing the grain orientation in one region of the specimen deformed along 45º RD-ND before deformation and after 1200 fatigue cycles are depicted in Figs. 5a and 5b, respectively. The test was stopped at zero load after the maximum compressive strain was attained. The grain orientation is random in this region and evidence of elongated twins (marked with arrows) can be found in the deformed specimen (Fig. 5b). The grain boundary misorientation angle at these twin boundaries was about 86° and they were acknowledged as tension twins (Fig. 5c). They were nucleated during the compressive part of the fatigue cycle and did not disappear by detwinning during unloading [16,44].

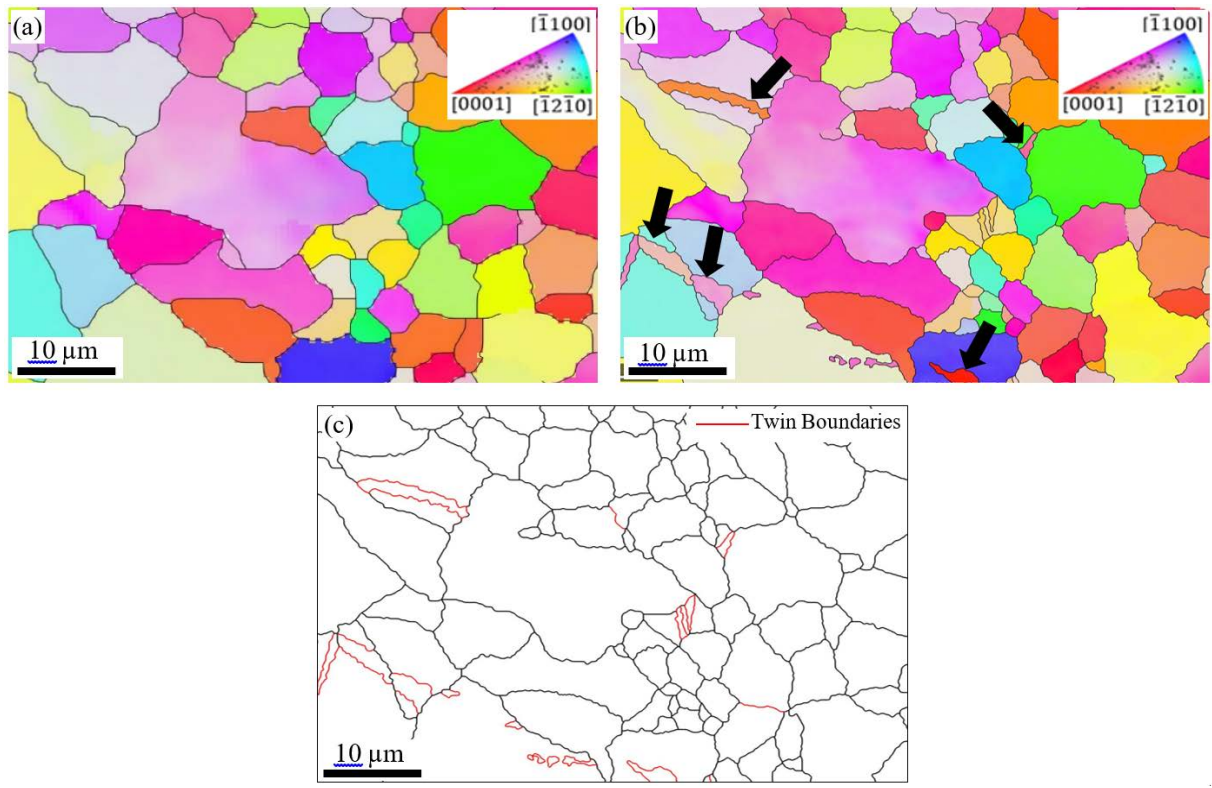

**Fig. 5.** EBSD maps of one region of the specimen deformed along 45º RD-ND. (a) Before deformation. (b) After 1200 fatigue cycles. (c) Grain boundaries and twin boundaries after deformation, the latter marked with red lines. The test was stopped at the maximum compressive strain. Tensile twins in (b) are marked with arrows.

Overall, the grain orientation on the surface of all the specimens was determined by EBSD and the orientation of the slip bands due to basal and pyramidal slip as well as the orientation of the six different tensile twin variants was ascertained in each grain by means of slip trace analysis following the methodology presented in [46,47]. They were compared with the slip band orientation and the elongated twins observed in the deformed specimens in the SEM using secondary electrons. An example of this analysis corresponding to different specimens deformed along 45º RD-ND at $\Delta\varepsilon/2 = 0.4\%$ and 2% is depicted in Figs. 6a and 6b, respectively. While the images in Figs 6c and 6d correspond to the specimen deformed along RD at $\Delta\varepsilon/2 = 0.4\%$ and Fig. 6e is taken from the specimen deformed along RD at $\Delta\varepsilon/2 = 2.0\%$. Only elongated bands parallel to tensile twins were found in the specimen deformed along 45º RD-ND at $\Delta\varepsilon/2 = 0.4\%$ (Fig. 6a), while elongated bands parallel to the twins and basal slip bands (marked respectively with blue and red lines) were found in the specimen deformed at $\Delta\varepsilon/2 = 2\%$. Obviously, deformation bands were not found in all the grains because twin boundaries did not reach the grain surface or because the orientation of the basal slip system was parallel to the grain surface and did not lead to any trace, but they clearly indicate the dominant deformation mechanisms in this orientation. The most likely explanation for the lack of basal slip traces in the specimen deformed along 45º RD-ND at $\Delta\varepsilon/2 = 0.4\%$ is that plastic deformation by basal slip was homogeneously distributed among many grains. As a result, the basal slip was not localized in slip bands in a few grains, that could be ascertained from secondary electron images in the SEM. Increasing $\Delta\varepsilon/2$ up to 2% led to the localization of basal slip in slip bands in many grains (Fig. 6b), as it has been reported previously in this alloy deformed in fatigue along 45º RD-ND at $\Delta\varepsilon/2 = 1\%$ [13]. These deformation mechanisms (basal slip and twinning) in the samples deformed along 45º RD-ND (representative of Mg alloys with weak basal texture) are in good agreement with those reported in the literature during fatigue of AZ31 Mg alloy along 45º RD-ND and of weakly textured Mg–3wt.% Y alloy deformed at $\Delta\varepsilon/2 = 1\%$ [11–13,17].

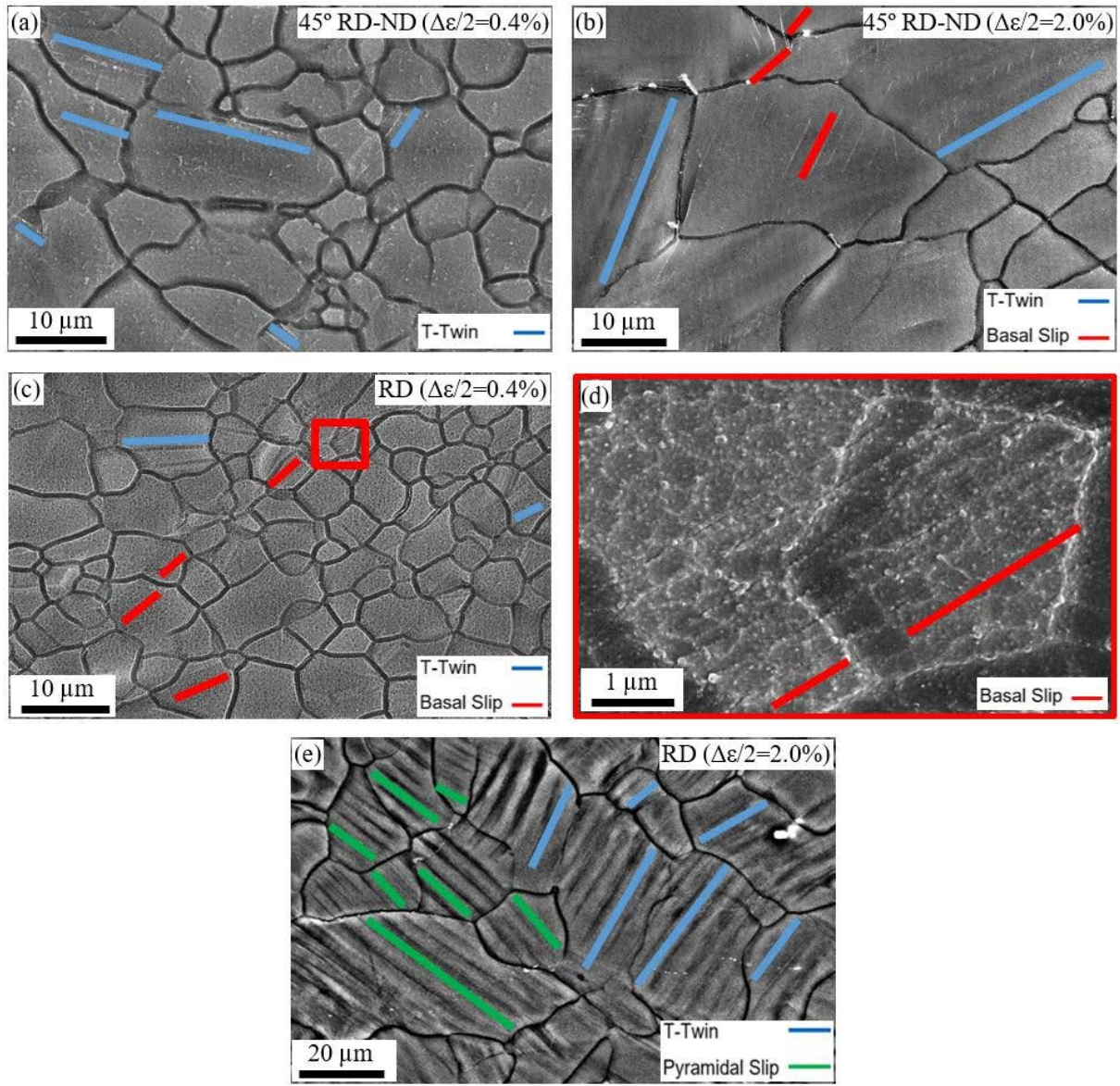

**Fig. 6.** Secondary electron image showing slip bands and elongated twins within the grains. (a) 1200 cycles at Δε/2 = 0.4% along 45º RD-ND. (b) 50 cycles at Δε/2 = 2.0% along 45º RD-ND. (c) and (d) 500 cycles at Δε/2 = 0.4% along RD. (e) 50 cycles at Δε/2 = 2.0% along RD. Following slip/twin trace analysis, those parallel to red lines were identified as basal slip bands, those parallel to the blue lines were tensile twins and those parallel to the green lines stand for pyramidal slip bands.

The analysis of the lateral surfaces of the specimen deformed along RD at Δε/2 = 0.4% led to the same conclusions and only elongated twins and basal slip bands were observed (Fig. 6c). The latter are always very fine and can be observed more clearly in Fig. 6d, which shows the area marked with a red rectangle in Fig. 6c at higher magnification. It should be noted that pyramidal slip traces were not observed in this specimen deformed at Δε/2 = 0.4%, indicating that the stresses during the tensile part of the fatigue cycle were not large enough to lead to the formation of pyramidal slips bands. Nevertheless, the pyramidal slip should be present in these samples to justify the asymmetry between the maximum and minimum stress in each fatigue cycle (Fig. 4b). Analysis of the fatigue deformation mechanism in an AZ31 Mg alloy by Shi et al. [13] also showed the activation of basal slip and twinning $\Delta\varepsilon/2 = 1\%$ along RD and ND. The analysis of the samples deformed along RD at $\Delta\varepsilon/2 = 2\%$ showed evidence of pyramidal slip bands that, together with tensile twins, were the dominant deformation mechanisms (Fig. 6e). Surprisingly, basal slip traces were not found in the slip trace analysis of the sample deformed along RD at $\Delta\varepsilon/2 = 2\%$, but it is obvious that basal slip should also be active at this cyclic strain semi-amplitude in this orientation. The most likely reason for the lack of basal slip traces is that area of the sample analyzed (Fig. 1a in [7]) had a very strong basal texture (stronger than the average one), and most of the basal planes were (almost) parallel to the RD-TD plane. Thus, the Burgers vectors of the basal dislocations were (almost) parallel to the surface, leaving no slip traces. On the contrary, regions in which the grains were suitably oriented for basal slip were found in the sample deformed along the RD at $\Delta\varepsilon/2 = 0.4\%$ (Fig. 1c), and basal slip traces were found in these grains.

In order to obtain quantitative information about the dominant deformation mechanisms, a titanic effort was carried out to search for slip traces and twins in several thousand grains for each orientation and strain amplitude (Table 2). The fraction of the grains with traces was much reduced in the specimen deformed along 45º RD-ND at $\Delta\varepsilon/2 = 0.4\%$ (3.8%) and increased when $\Delta\varepsilon/2 = 2\%$ and also when the samples were oriented along the RD.

According to the slip trace analysis, plastic deformation was accommodated by the formation of tensile twins (followed by detwinning), which were found in 3.8% of the grains analyzed (Table 2) in the specimen deformed along 45º RD-ND at $\Delta\varepsilon/2 = 0.4\%$. Basal slip traces were not found in this specimen. However, it is important to mention that basal slip should also be a dominant deformation mechanism because most grains were suitable oriented for basal slip [11,13,16,28]. The number of grains with slip traces increased to 16% in the specimen deformed along 45º RD-ND at $\Delta\varepsilon/2 = 2\%$, and approximately one-half of them showed elongated twins. The other half showed basal slip traces, which appeared because slip localization was triggered at higher cyclic strain semi-amplitude.

**Table 2.** Deformation mechanisms from slip/twin trace analysis.

| Orientation | 45º RD-ND | | RD | |
|---|---|---|---|---|
| $\Delta\varepsilon/2$ (%) | 0.4 | 2.0 | 0.4 | 2.0 |
| Number of grains analyzed | 4515 | 1544 | 2348 | 2100 |
| Number of grains with traces | 171 | 248 | 170 | 538 |
| Elongated tensile twins | 100% | 52.4% | 61.2% | 18.4% |
| Basal slip | - | 47.6% | 38.8% | 0.2% |
| Pyramidal slip | - | - | - | 72.3% |
| Pyramidal slip & tensile twins | - | - | - | 9.1% |

Elongated twins and basal slip bands (with an approximation proportion of 60/40) were found in 7.2% of the grains of the specimen deformed along RD at $\Delta\varepsilon/2 = 0.4\%$ (Table 2), and they stand for the dominant deformation mechanisms according to slip trace analysis. The presence of basal slip bands can be explained because the fraction of grains suitable oriented for basal slip is very limited due to the strong basal texture and, thus, the deformation has to be accommodated in fewer grains, which take up more plastic deformation. In fact, basal slip traces were often found in elongated clusters of small grains suitable oriented for basal slip. The orientation of the cluster of grains was parallel to the basal slip traces (Fig. 6c), and these regions in which basal slip was present in several contiguous grains were important from the viewpoint of crack nucleation, as it will be shown in the following section. The asymmetry in the maximum stresses in tension and compression seems to indicate that pyramidal slip was also active in this specimen, but very likely, its magnitude was too low to promote the formation of pyramidal slip bands on the surface. Shi et al. [13] did not detect pyramidal slip traces in fatigue of AZ31 Mg alloy at $\Delta\varepsilon/2 = 1\%$ along RD and ND. Nevertheless, the pyramidal slip was, together with tensile twinning, the dominant deformation mechanism according to slip trace analysis when $\Delta\varepsilon/2 = 2\%$ along RD (Fig. 6e). They were found in 25.6% of the grains, and pyramidal slip bands were dominant (Table 2). A small fraction of the grains which presented evidence of deformation included both pyramidal slip band and elongated twins (9.1%), indicating that both mechanisms can appear for given grain orientation. These results are in agreement with the large tension-compression anisotropy in the cyclic stress-strain curves, which was due to the activation of pyramidal slip during the tensile part of the cycle as basal slip was not able to accommodate the plastic deformation because of the strong basal texture. Overall, the conclusions of slip trace analysis for the fatigue deformation mechanisms are in agreement with those reported in other studies of Mg alloys deformed in different directions and with different strain amplitudes [11–13,16,25].

3.4 Fatigue crack initiation mechanisms

Cracks nucleated during the fatigue tests were identified in the same regions used to explore the deformation mechanisms in each specimen. The large size of the area analyzed allowed to include between 150 to 500 cracks in each specimen (Table 3), leading to statistically relevant results. Most of the cracks (73%) in the specimen deformed along 45º RD-ND at Δε/2 = 0.4% were nucleated at the twin/matrix boundary (parallel to the elongated twins), and one example is depicted in Figs. 7a and 7b. The remaining 27% of cracks were nucleated at grain boundaries, including triple points, and an example is depicted in Fig. 7c. In some cases, transgranular cracks nucleated at twins propagated along the grain boundaries, as shown in Fig. 7d. No other fatigue crack nucleation sites were observed in this specimen.

**Table 3.** Fatigue crack initiation locations as a function of specimen orientation and cyclic strain semi-amplitude. IC stands for intergranular cracking and TC for transgranular cracking.

| Orientation | 45º RD-ND | | RD | |
|---|---|---|---|---|
| Δε/2 (%) | 0.4 | 2.0 | 0.4 | 2.0 |
| Number of cracks | 212 | 289 | 152 | 496 |
| Grain boundary crack (IC) | 26.9% | 67.1% | 62.5% | 57.2% |
| Basal slip band crack (TC) | - | 4.5% | 32.9% | - |
| Tensile twin boundary crack (TC) | 73.1% | 14.5% | 4.6% | 8.1% |
| Pyramidal slip band crack (TC) | - | - | - | 33.9% |
| Basal slip band crack within tensile twin (TC) | - | 13.9% | - | - |

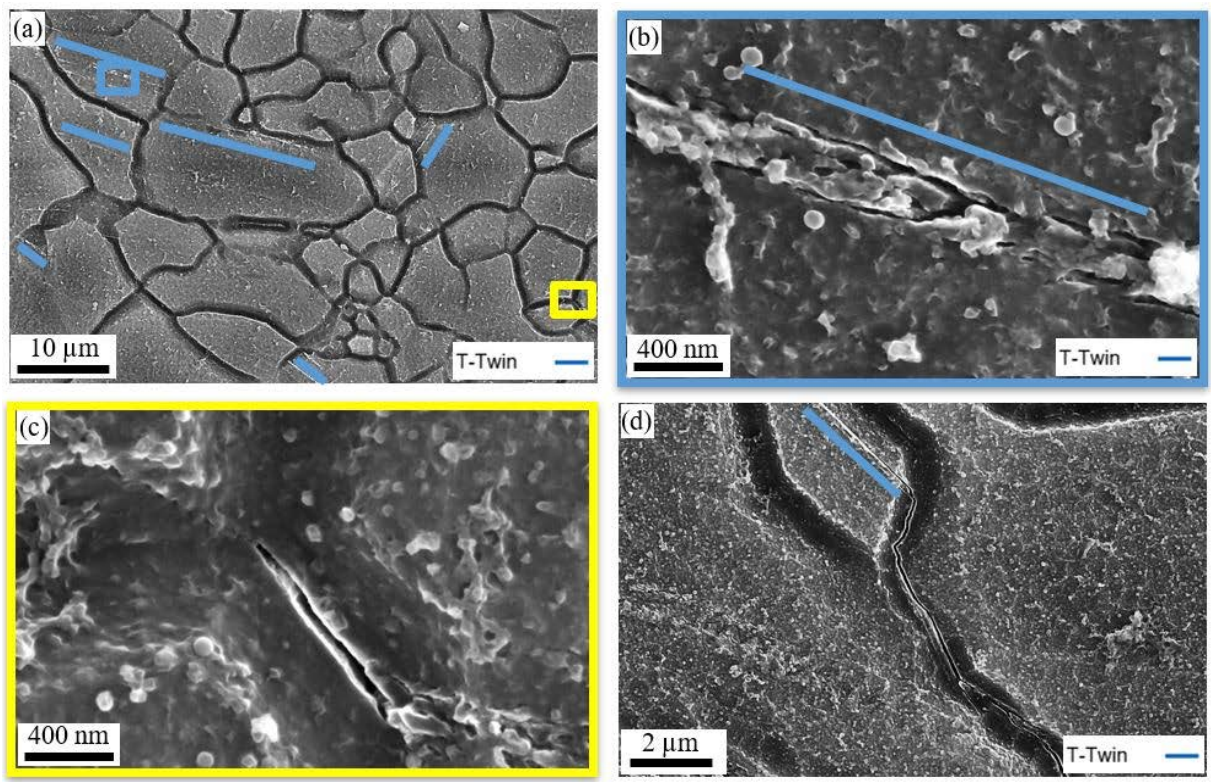

**Fig. 7.** (a) Fatigue crack initiation sites in the rolled AZ31B-O sample after 1200 fatigue cycles at Δε/2 = 0.4% along 45º RD-ND. (a) Crack parallel to twin. (b) Higher magnification of the region within the blue rectangle in (a). (c) Higher magnification of a grain boundary crack at a triple point within the yellow rectangle in (a). (d) Propagation along the grain boundary of a crack nucleated parallel to a twin.

Cracks nucleated parallel to the twins (Figs. 8a and 8b) as well as at triple points (Fig. 8c) were also observed for this orientation when Δε/2 = 2.0%, but the proportion has been inverted, and crack initiation at triple points was dominant (67%) with respect to cracks initiation parallel to twin (14.5%) (Table 3). A higher fraction of grain boundary cracks has also been reported in the literature in the fatigue of AZ31 Mg alloy along 45º RD-ND at Δε/2 = 1% (60.5% of the cracks were grain boundary cracks) [13]. In addition, clusters of short cracks parallel to the basal plane orientation were found within the twins, which are depicted in Figs. 8d and 8e. They were reported in the investigations [32] and are likely associated with the activation of basal slip within the twins during the tensile or compressive part of the fatigue cycle (twinned regions are suitable oriented for basal slip). Nevertheless, the fraction of these clusters is limited (13.9%), and they never propagated. Thus, they do not seem to be critical from the viewpoint of fatigue failure. Finally, a few cracks nucleated along basal slip bands (4.5%) were also found.

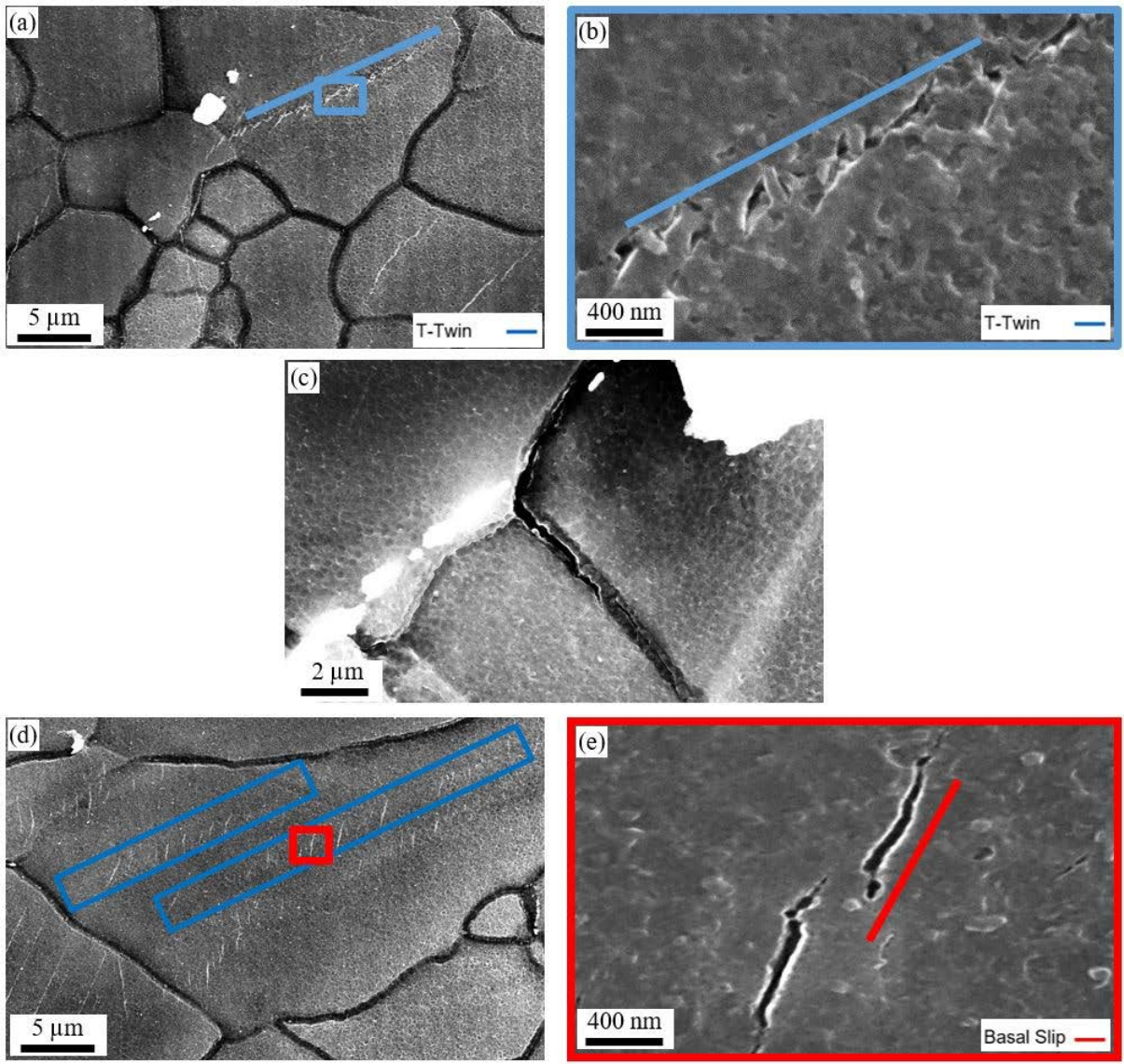

**Fig. 8.** Fatigue crack initiation sites in the rolled AZ31B-O sample after 50 fatigue cycles at Δε/2 = 2.0% along 45º RD-ND. (a) Crack parallel to a twin. (b) Higher magnification of the region within the blue rectangle in (a). (c) Intergranular crack at a grain boundary triple junction. (d) Short cracks parallel to the traces of basal planes within the twins. (e) Higher magnification of the region within the red rectangle shows the short cracks parallel to the traces of basal planes within the twins.

The fatigue crack nucleation sites along the 45º RD-ND analysis pointed out two different mechanisms: transgranular cracks were nucleated parallel to the twins or basal slip, and intergranular cracks appeared at grain boundaries and triple points. The twin cracks dominate at low cyclic strain semi-amplitudes (Δε/2 = 0.4%) and the grain boundary cracks dominate at high cyclic strain semi-amplitudes (Δε/2 = 2%). Very likely, the activation of basal slip bands at Δε/2 = 2% (Fig. 6b) led to the development of stress concentrations at grain boundaries and triple points, facilitating crack nucleation. These types of cracks were in good agreement with those reported in previous investigations on the fatigue crack nucleation of weakly textured Mg alloys [13,37]. The probability of cracking as a function of the grain size for the different types of crack initiation sites found in these specimens is plotted in Figs. 9a and 9b for Δε/2 = 0.4% and 2%, respectively. In the case of grain boundaries, the grain size was the average of the adjacent grains. These results clearly indicate that the preferential crack initiation sites are grain boundaries and triple points between small grains and transgranular cracks parallel to the twins in large grains. It should be noted that the probability of nucleating a fatigue crack is very high in the latter (above 40%), and these cracks are very long because they encompass the full width of the large grain. Thus, large grains were oriented for twinning seem to be the preferential location for the nucleation of fatigue cracks leading to failure in specimens in which basal slip and tensile twinning are the dominant deformation mechanisms. Finally, the effect of grain boundary misorientation on the probability of nucleating a crack at the grain boundaries was explored, and the results are plotted in Fig. 9c. This factor did not play any role in the nucleation of fatigue cracks at grain boundaries in the specimens deformed along 45º RD-ND at both Δε/2 = 0.4% and 2%, although the crack initiation probability increases with Δε/2.

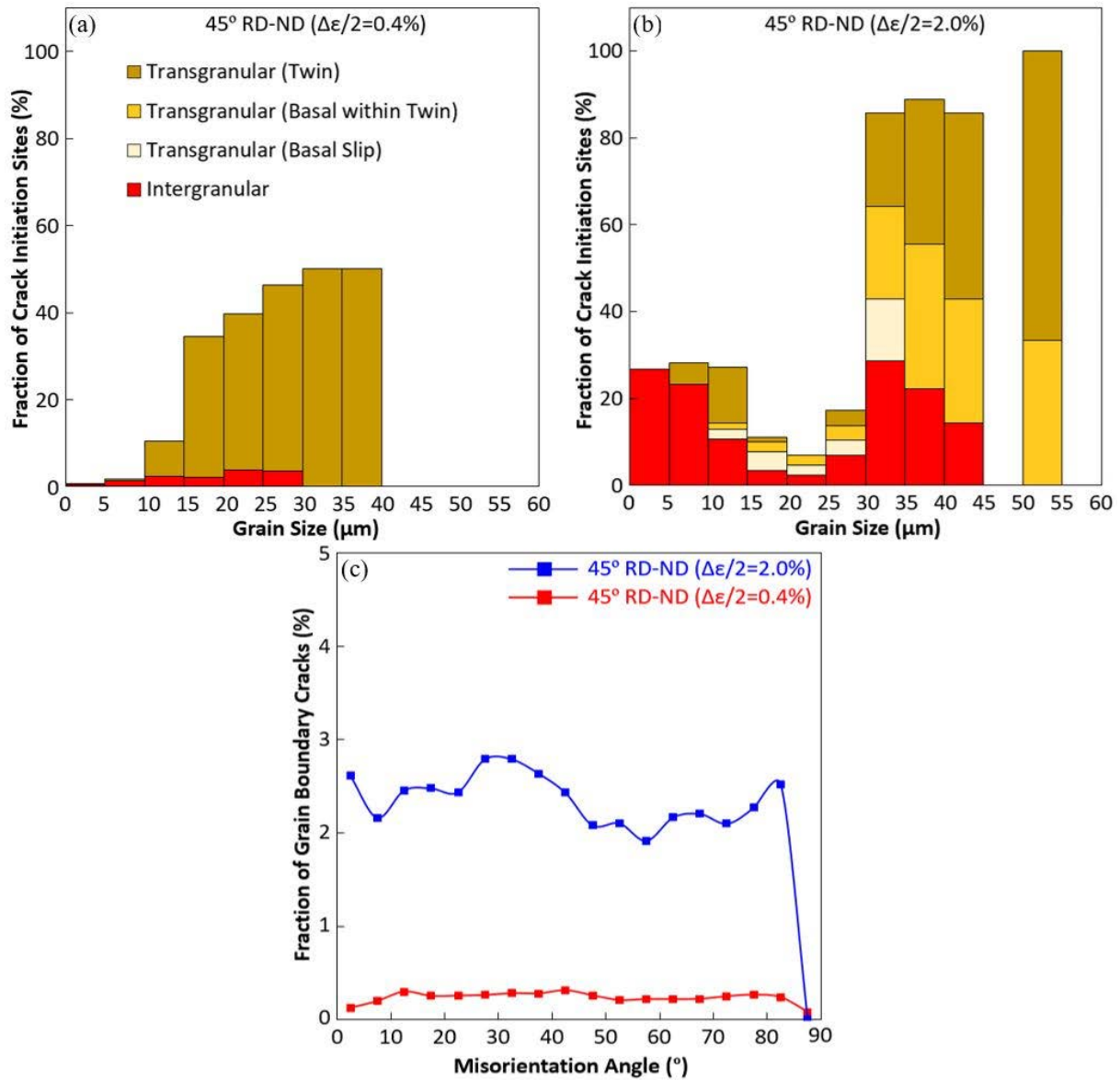

**Fig. 9.** (a) Fraction of crack initiation sites as a function of grain size for each crack initiation mechanism in the specimen deformed along 45º RD-ND at Δε/2 = 0.4%. (b) *Idem* at Δε/2 = 2%. (c) Fraction of intergranular cracks as a function of the grain boundary misorientation angle in the specimens deformed along 45º RD-ND.

The crack nucleation sites in the specimens deformed along RD at Δε/2 = 0.4% are depicted in Fig. 10. The most important one (62.5%, Table 3) was intergranular cracking that developed at triple points (yellow rectangle in Figs. 10a and 10c). The second fatigue crack initiation mechanism (32.9%, Table 3) was associated with cracks parallel to basal slip bands (red rectangle in Fig. 10a and 10b), which developed in clusters of grains suitably oriented for basal slip. They often propagated through the grain boundaries to link several grains. A few cracks parallel to the elongated twin boundaries in large grains were also found (4.5%, Table 3) (Fig. 10d and 10e).

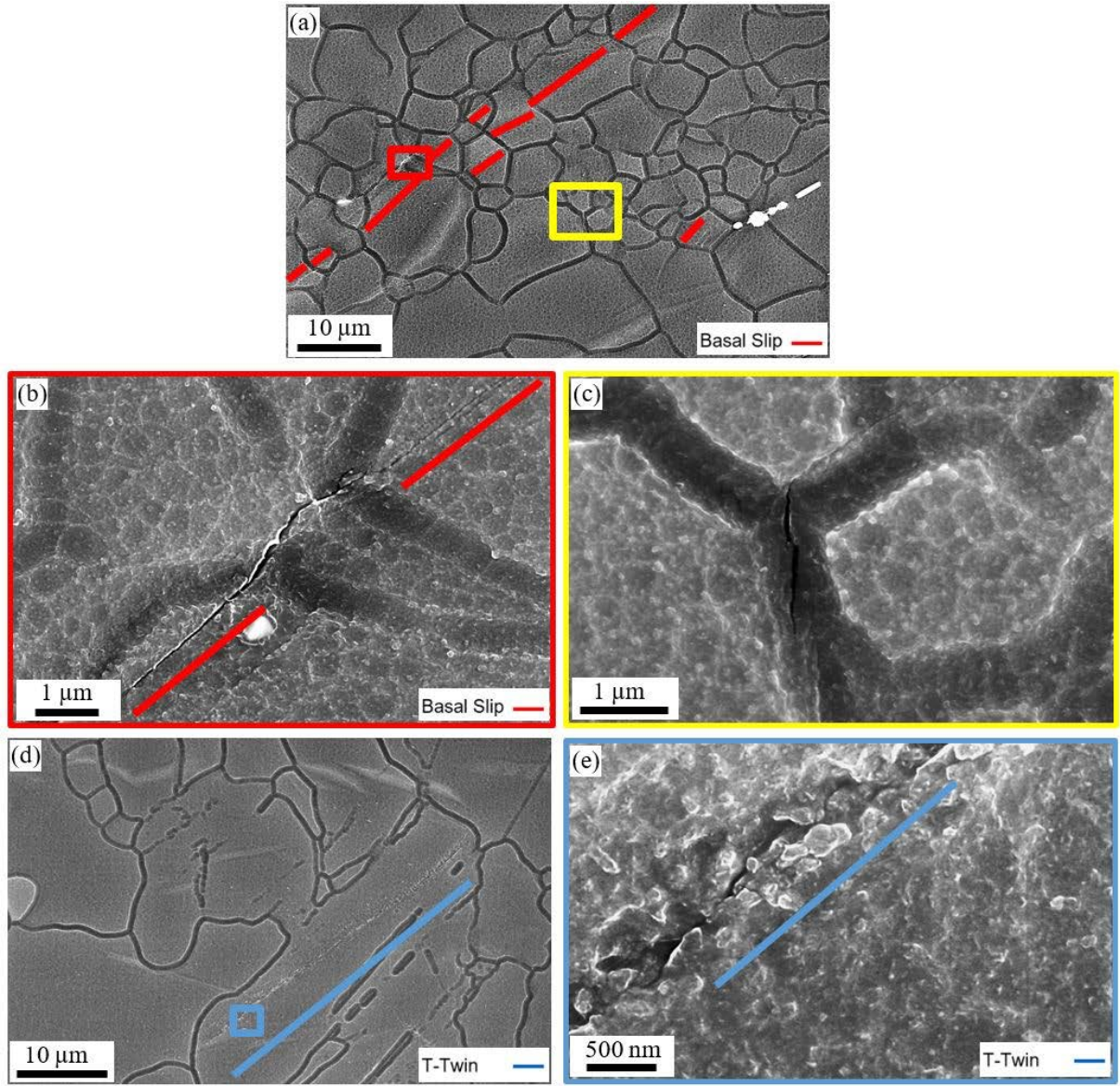

**Fig. 10.** Fatigue crack initiation sites in the rolled AZ31B-O sample after 50 fatigue cycles at Δε/2 = 0.4% along RD. (a) Intergranular cracks near a cluster of grains with almost parallel basal slip traces (yellow rectangle) and at a triple point (red rectangle). (b) Higher magnification of the yellow rectangle in (a). (c) Higher magnification of the red rectangle in (a). (d) Crack parallel to a twin. (e) Higher magnification of the region within the blue rectangle in (d).

When Δε/2 increased to 2%, the dominant fatigue crack initiation sites were cracks parallel to the pyramidal slip bands (Fig. 11a and 11b) (33,9%, Table 3) and grain boundary cracks (57.2%, Table 3). An interesting example of a grain boundary crack is found in Figs 11c and 11d, in which the crack has appeared around a small grain that is fully twinned as a result of the eigenstrain associated with twinning. In addition, cracks parallel to the twin boundaries were also found (8.1%, Table 3), and they were linked to large grains.

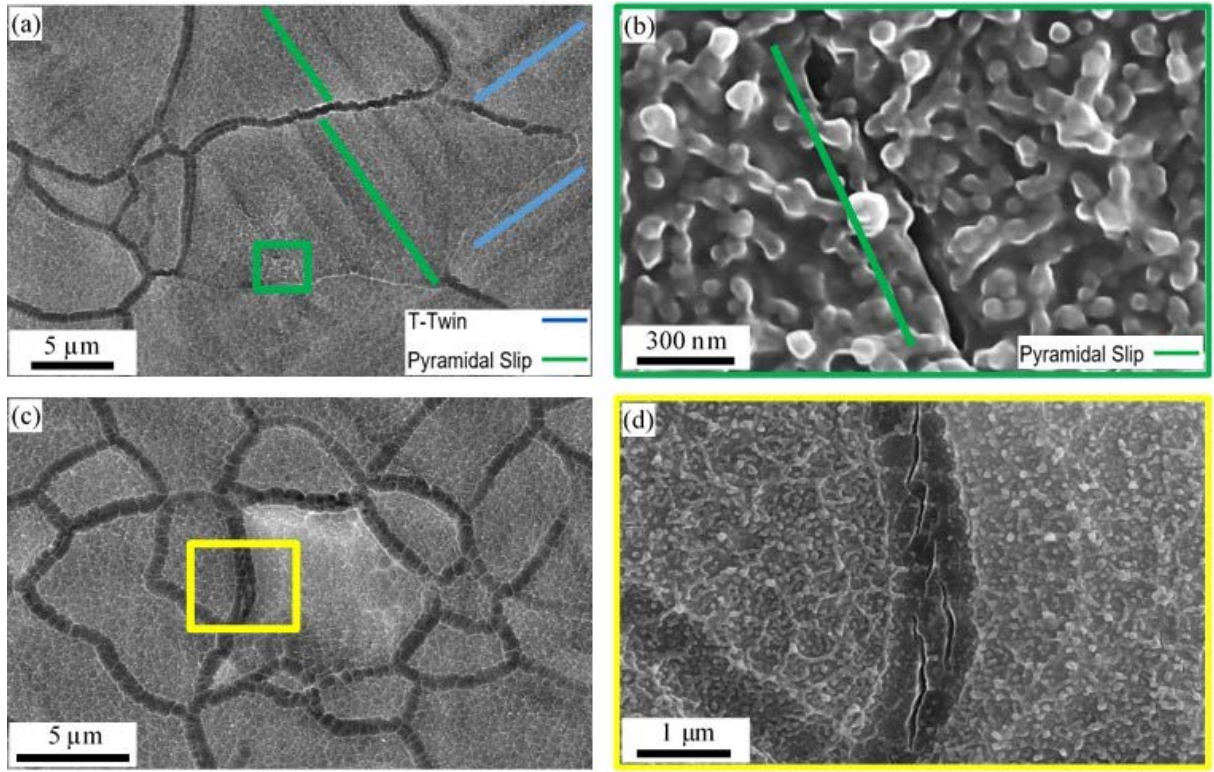

**Fig. 11.** Fatigue crack initiation sites in the rolled AZ31B-O sample after 50 fatigue cycles at $\Delta\varepsilon/2$ = 2.0% along RD. (a) Crack parallel to pyramidal slip band. (b) Higher magnification of the region within the green rectangle in (a). (c) Intergranular crack at a grain boundary around a fully-twinned small grain. (d) Higher magnification of the region within the yellow rectangle in (c).

The analysis of fatigue crack nucleation sites along RD showed two different mechanisms: intergranular cracks at grain boundaries and triple points and transgranular cracks parallel to the basal slip bands or to pyramidal slip bands. The former transgranular cracks are dominant at low cyclic strain semi-amplitudes ($\Delta\varepsilon/2$ = 0.4%) and the latter at high cyclic strain semi-amplitudes ($\Delta\varepsilon/2$ = 2%) because pyramidal slip is very limited at $\Delta\varepsilon/2$ = 0.4%. In addition, a few transgranular cracks parallel to the twin boundaries were observed at both cyclic strain semi-amplitudes. The probability of cracking as a function of the grain size for the different types of crack initiation sites found in these specimens is plotted in Figs. 12a and 12b for $\Delta\varepsilon/2$ = 0.4% and 2%, respectively. They show that the probability of cracking was very high in large grains (> 30 μm). These transgranular cracks developed parallel to the twin boundaries at $\Delta\varepsilon/2$ = 0.4% and along the pyramidal slip bands at $\Delta\varepsilon/2$ = 2.0%. They encompassed the full width of the grain (see, for instance, Figs. 10d and 11a) and were likely to propagate until failure. In addition, intergranular cracking was dominant around small grains, together with transgranular cracks associated with twin boundaries or basal/pyramidal slip bands. In general, these intergranular or transgranular cracks around or through small grains were short and, thus, less critical from the viewpoint of propagation until failure. The only exception was transgranular cracks parallel to basal slip bands in clusters of grains suitably oriented for basal slip (Fig. 10a and 10b). They grew rapidly by linking up cracks parallel to basal slip bands in several grains and can also limit the fatigue strength of the material. Finally, the effect of grain boundary misorientation on the fatigue crack initiation probability at grain boundaries is plotted in Fig. 12c. While there is no correlation between both factors at $\Delta\varepsilon/2 = 0.4\%$, grain boundary cracks were associated with high angle grain boundaries (>40º) at $\Delta\varepsilon/2 = 2.0\%$. A clear explanation for this phenomenon is not available, although it may be related to the lack of slip transfer at high-angle grain boundaries in the presence of pyramidal slip.

Thus, the experimental observations presented above provide clear indications about the microstructural factors that control the nucleation of the most damaging fatigue cracks that determine the fatigue behavior of Mg alloys. In the case of Mg alloys with low texture and/or with most of the grains suitable oriented to activate the softest deformation mechanisms (basal slip and tensile twinning), the most damaging fatigue cracks nucleate parallel to twins in large grains. Grain boundary cracks also appear during deformation, particularly around small grains, and the intergranular fracture probability increases with the cyclic strain semi-amplitude because of the higher stress concentrations associated with the localization of slip in the surrounding grains. However, these cracks are shorter than those nucleated parallel to twins in large grains; thus, the latter is more critical from the viewpoint of fatigue failure.

The fatigue crack nucleation mechanisms change in the case of textured Mg alloys deformed parallel or perpendicular to the basal plane, which leads to a large anisotropy in the cyclic stress-strain curve and to the activation of twinning in the tensile or compressive part of the fatigue cycle and to the activation of pyramidal slip in the other part of the fatigue cycle. Fatigue cracks are nucleated along tensile twin boundaries in large grains at low cyclic strain semi-amplitudes and along pyramidal slip bands in large grains at high cyclic strain semi-amplitudes. In both cases, intergranular cracks appear at grain boundaries around small grains, but they are short and not as critical with respect to fatigue failure as those nucleated along pyramidal slip bands or twins in large grains.

These results indicate that large grains are very detrimental from the viewpoint of the fatigue life of Mg alloys, as well as the presence of clusters of grains suitably oriented for basal slip, which tend to localize the deformation. In addition, intergranular cracking around small grains often occurs during cyclic deformation due to the stress concentrations associated with the plastic anisotropy of the Mg crystals. Thus, fatigue indicator parameters based on the accumulated plastic strain associated with basal, pyramidal slip, and twinning/detwinning in each cycle seem to be appropriate to predict the fatigue life of Mg alloys [9,41].

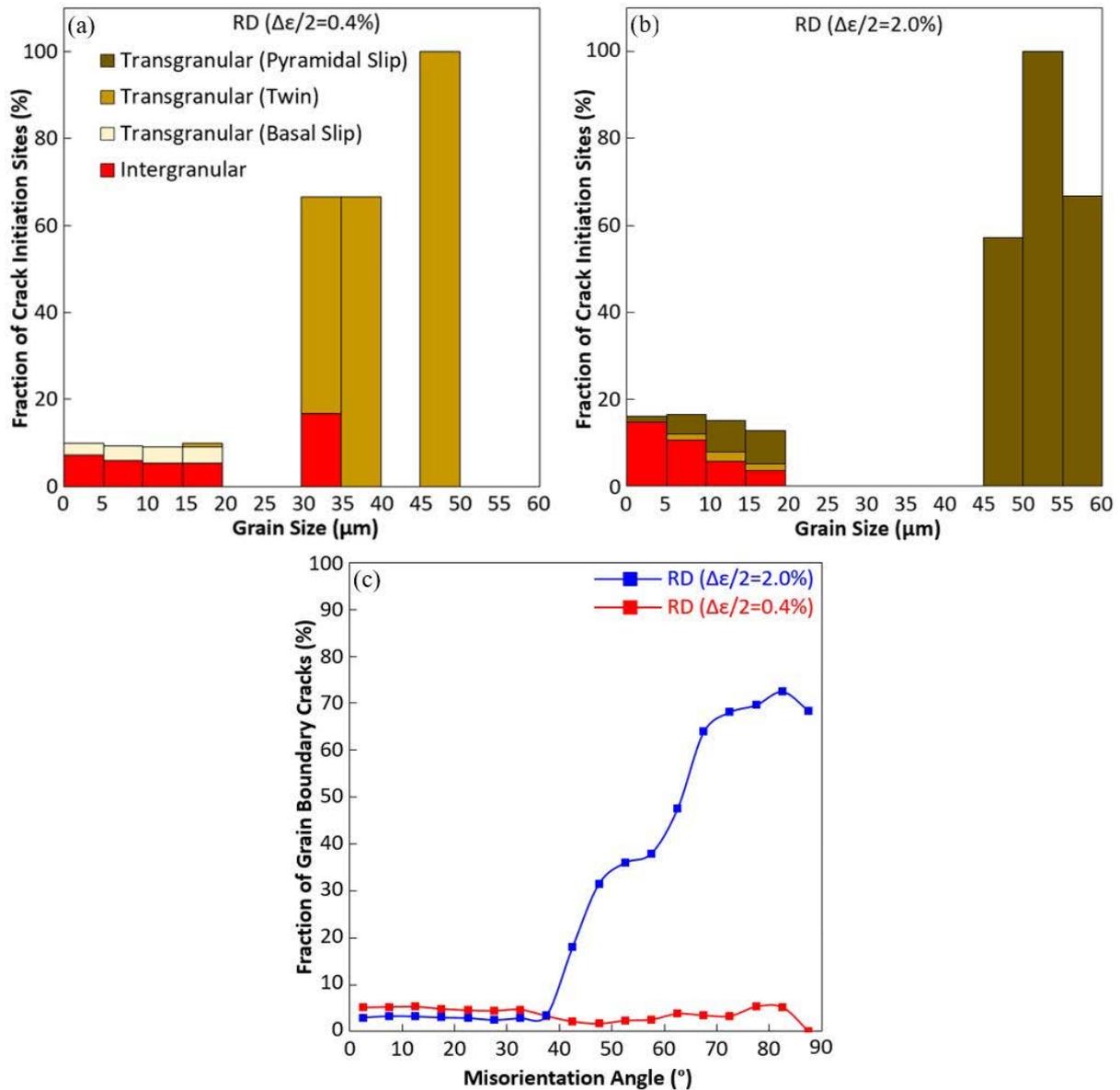

**Fig. 12.** (a) Fraction of crack initiation sites as a function of grain size for each crack initiation mechanism in the specimen deformed along RD at $\Delta\varepsilon/2 = 0.4\%$. (b) *Idem* at $\Delta\varepsilon/2 = 2\%$. (c) Fraction of intergranular cracks as a function of the grain boundary misorientation angle in the specimens deformed along RD.

## 4. Conclusions

The deformation and fatigue crack nucleation mechanisms under fully-reversed, strain-controlled cyclic deformation were ascertained in a textured AZ31B-O alloy at $\Delta\varepsilon/2 = 0.4\%$ and 2.0% along the rolling direction and at 45° between normal and rolling directions by means of slip trace analysis and electron microscopy. Between 1500 and 4500 grains were analyzed for each orientation and cyclic strain semi-amplitude to obtain statistically significant results. Specimens oriented at 45º between rolling and normal directions were suitably oriented for basal slip and twining/detwinning, which were the dominant deformation mechanisms. They are representative of Mg alloys with low texture. The most damaging fatigue cracks were initiated parallel to twins in large grains. Grain boundary cracks, particularly around small grains, were also found, and the intergranular fracture probability increased with the cyclic strain semi-amplitude, but these cracks are shorter and less critical from the viewpoint of fatigue failure.

The main deformation mechanisms in the specimens oriented along the rolling direction were tensile twinning/detwinning and basal slip at low cyclic strain semi-amplitudes and tensile twinning/detwinning and pyramidal slip at high cyclic strain semi-amplitudes. Large cracks were nucleated along twins and pyramidal slip bands at low and high cyclic strain semi-amplitudes, respectively, in large grains. In addition, localization of deformation in clusters of small grains suitable oriented for basal slip also led to the nucleation of large fatigue cracks. Grain boundary cracks, particularly around small grains, were also found but they were shorter and less critical from the viewpoint of fatigue failure

**Declaration of Competing Interest**

The authors declare that they have no known competing financial interests or personal relationships that could have appeared to influence the work reported in this paper.

**Acknowledgments**

This investigation was supported by the European Union Horizon 2020 research and innovation programme (Marie Sklodowska-Curie Individual Fellowships, Grant Agreement 795658) and the Comunidad de Madrid Talento-Mod1 programme (Grant Agreement PR-00096). Additional support by the HexaGB project of the Spanish Ministry of Science (reference RTI2018-098245) is also gratefully acknowledged.

**References**

[1]   S. Suresh, Fatigue of Materials, Cambridge University Press, 1998. https://doi.org/10.1017/CBO9780511806575.

[2]   M.D. Sangid, The physics of fatigue crack initiation, Int. J. Fatigue. 57 (2013) 58–72. https://doi.org/10.1016/j.ijfatigue.2012.10.009.

[3]   J.C. Tucker, A.R. Cerrone, A.R. Ingraffea, A.D. Rollett, Crystal plasticity finite element analysis for René88DT statistical volume element generation, Model. Simul. Mater. Sci.

Eng. 23 (2015) 35003. https://doi.org/10.1088/0965-0393/23/3/035003.

[4] J.C. Stinville, W.C. Lenthe, M.P. Echlin, P.G. Callahan, D. Texier, T.M. Pollock, Microstructural statistics for fatigue crack initiation in polycrystalline nickel-base superalloys, Int. J. Fract. 208 (2017) 221–240. https://doi.org/10.1007/s10704-017-0241-z.

[5] P.D. Littlewood, A.J. Wilkinson, Local deformation patterns in Ti-6Al-4V under tensile, fatigue and dwell fatigue loading, Int. J. Fatigue. 43 (2012) 111–119. https://doi.org/10.1016/j.ijfatigue.2012.03.001.

[6] F.P.E. Dunne, A. Walker, D. Rugg, A systematic study of hcp crystal orientation and morphology effects in polycrystal deformation and fatigue, Proc. R. Soc. A Math. Phys. Eng. Sci. 463 (2007) 1467–1489. https://doi.org/10.1098/rspa.2007.1833.

[7] A. Jamali, A. Ma, J. Llorca, Scripta Materialia Influence of grain size and grain boundary misorientation on the fatigue crack initiation mechanisms of textured AZ31 Mg alloy, Scr. Mater. 207 (2022) 114304. https://doi.org/10.1016/j.scriptamat.2021.114304.

[8] Y. Nakai, M. Saka, H. Yoshida, K. Asayama, S. Kikuchi, Fatigue crack initiation site and propagation paths in high-cycle fatigue of magnesium alloy AZ31, Int. J. Fatigue. 123 (2019) 248–254. https://doi.org/10.1016/j.ijfatigue.2019.02.024.

[9] F. Briffod, T. Shiraiwa, M. Enoki, Numerical investigation of the influence of twinning/detwinning on fatigue crack initiation in AZ31 magnesium alloy, Mater. Sci. Eng. A. 753 (2019) 79–90. https://doi.org/10.1016/j.msea.2019.03.030.

[10] A.D. Murphy-Leonard, D.C. Pagan, A. Beaudoin, M.P. Miller, J.E. Allison, Quantification of cyclic twinning-detwinning behavior during low-cycle fatigue of pure magnesium using high energy X-ray diffraction, Int. J. Fatigue. 125 (2019) 314–323. https://doi.org/10.1016/j.ijfatigue.2019.04.011.

[11] T. Li, J. Zheng, H. Shou, R. Shi, Y. Zhang, G. Wen, K. Huang, D. Yin, J. Rao, The deformation modes and transferability during low-cycle fatigue of Mg and Mg–3Y alloy, Mater. Sci. Eng. A. 839 (2022) 142838. https://doi.org/10.1016/j.msea.2022.142838.

[12] T. Li, J. Rao, J. Zheng, D. Yin, H. Shou, Y. Zhang, Anisotropic cyclic deformation behavior of an extruded Mg-3Y alloy sheet with rare earth texture, J. Magnes. Alloy. (2022). https://doi.org/10.1016/j.jma.2022.05.010.

[13] R. Shi, J. Zheng, T. Li, H. Shou, D. Yin, J. Rao, Quantitative analysis of the deformation modes and cracking modes during low-cycle fatigue of a rolled AZ31B magnesium alloy: The influence of texture, Mater. Sci. Eng. A. 844 (2022) 143103. https://doi.org/10.1016/j.msea.2022.143103.

[14] L. Wu, A. Jain, D.W. Brown, G.M. Stoica, S.R. Agnew, B. Clausen, D.E. Fielden, P.K. Liaw, Twinning-detwinning behavior during the strain-controlled low-cycle fatigue testing of a wrought magnesium alloy, ZK60A, Acta Mater. 56 (2008) 688–695. https://doi.org/10.1016/j.actamat.2007.10.030.

[15] L. Wu, S.R. Agnew, D.W. Brown, G.M. Stoica, B. Clausen, A. Jain, D.E. Fielden, P.K. Liaw, Internal stress relaxation and load redistribution during the twinning – detwinning-dominated cyclic deformation of a wrought magnesium alloy , ZK60A, Acta

Mater. 56 (2008) 3699–3707. https://doi.org/10.1016/j.actamat.2008.04.006.

[16] Y. Wang, D. Culbertson, Y. Jiang, An experimental study of anisotropic fatigue behavior of rolled AZ31B magnesium alloy, Mater. Des. 186 (2020) 108266. https://doi.org/10.1016/j.matdes.2019.108266.

[17] Y.C. Deng, Z.J. Huang, T.J. Li, D.D. Yin, J. Zheng, Quantitative Investigation on the Slip / Twinning Activity and Cracking Behavior During Low-Cycle Fatigue of an Extruded Mg-3Y Sheet, Metall. Mater. Trans. A. 52 (2021) 332–349. https://doi.org/10.1007/s11661-020-06083-7.

[18] C.J. Geng, B.L. Wu, X.H. Du, Y.D. Wang, Y.D. Zhang, F. Wagner, C. Esling, Stress-strain response of textured AZ31B magnesium alloy under uniaxial tension at the different strain rates, Mater. Sci. Eng. A. 559 (2013) 307–313. https://doi.org/10.1016/j.msea.2012.08.103.

[19] Q. Yu, J. Zhang, Y. Jiang, Direct observation of twinning-detwinning-retwinning on magnesium single crystal subjected to strain-controlled cyclic tension-compression in [0 0 0 1] direction, Philos. Mag. Lett. 91 (2011) 757–765. https://doi.org/10.1080/09500839.2011.617713.

[20] M. Yaghoobi, J.E. Allison, V. Sundararaghavan, Multiscale modeling of twinning and detwinning behavior of HCP polycrystals, Int. J. Plast. 127 (2020) 102653. https://doi.org/10.1016/j.ijplas.2019.102653.

[21] H. Zhang, A. Jérusalem, E. Salvati, C. Papadaki, K.S. Fong, X. Song, A.M. Korsunsky, Multi-scale mechanisms of twinning-detwinning in magnesium alloy AZ31B simulated by crystal plasticity modeling and validated via in situ synchrotron XRD and in situ SEM-EBSD, Int. J. Plast. 119 (2019) 43–56. https://doi.org/10.1016/j.ijplas.2019.02.018.

[22] D.D. Yin, C.J. Boehlert, L.J. Long, G.H. Huang, H. Zhou, J. Zheng, Q.D. Wang, D. Wang, Tension-compression asymmetry and the underlying slip/twinning activity in extruded Mg-Y sheets, Int. J. Plast. 136 (2021) 102878. https://doi.org/10.1016/j.ijplas.2020.102878.

[23] E. Dogan, I. Karaman, G. Ayoub, G. Kridli, Reduction in tension-compression asymmetry via grain refinement and texture design in Mg-3Al-1Zn sheets, Mater. Sci. Eng. A. 610 (2014) 220–227. https://doi.org/10.1016/j.msea.2014.04.112.

[24] Z. Zachariah, S.S. V. Tatiparti, S.K. Mishra, N. Ramakrishnan, U. Ramamurty, Tension-compression asymmetry in an extruded Mg alloy AM30: Temperature and strain rate effects, Mater. Sci. Eng. A. 572 (2013) 8–18. https://doi.org/10.1016/j.msea.2013.02.023.

[25] S.H. Park, S.G. Hong, W. Bang, C.S. Lee, Effect of anisotropy on the low-cycle fatigue behavior of rolled AZ31 magnesium alloy, Mater. Sci. Eng. A. 527 (2010) 417–423. https://doi.org/10.1016/j.msea.2009.08.044.

[26] S.H. Park, S.G. Hong, J. Yoon, C.S. Lee, Influence of loading direction on the anisotropic fatigue properties of rolled magnesium alloy, Int. J. Fatigue. 87 (2016) 210–215. https://doi.org/10.1016/j.ijfatigue.2016.01.026.

[27] S. Dong, Y. Jiang, J. Dong, F. Wang, W. Ding, Cyclic deformation and fatigue of extruded ZK60 magnesium alloy with aging effects, Mater. Sci. Eng. A. 615 (2014)

262–272. https://doi.org/10.1016/j.msea.2014.07.074.

[28] Y. Xiong, Y. Jiang, Cyclic deformation and fatigue of rolled AZ80 magnesium alloy along different material orientations, Mater. Sci. Eng. A. 677 (2016) 58–67. https://doi.org/10.1016/j.msea.2016.09.031.

[29] Y. Xiong, Q. Yu, Y. Jiang, An experimental study of cyclic plastic deformation of extruded ZK60 magnesium alloy under uniaxial loading at room temperature, Int. J. Plast. 53 (2014) 107–124. https://doi.org/10.1016/j.ijplas.2013.07.008.

[30] F. Lv, F. Yang, Q.Q. Duan, Y.S. Yang, S.D. Wu, S.X. Li, Z.F. Zhang, Fatigue properties of rolled magnesium alloy (AZ31) sheet: Influence of specimen orientation, Int. J. Fatigue. 33 (2011) 672–682. https://doi.org/10.1016/j.ijfatigue.2010.10.013.

[31] Q. Li, Q. Yu, J. Zhang, Y. Jiang, Effect of strain amplitude on tension-compression fatigue behavior of extruded Mg6Al1ZnA magnesium alloy, Scr. Mater. 62 (2010) 778–781. https://doi.org/10.1016/j.scriptamat.2010.01.052.

[32] B. Wen, F. Wang, L. Jin, J. Dong, Fatigue damage development in extruded Mg-3Al-Zn magnesium alloy, Mater. Sci. Eng. A. 667 (2016) 171–178. https://doi.org/10.1016/j.msea.2016.05.009.

[33] Y. Xiong, Q. Yu, Y. Jiang, Cyclic deformation and fatigue of extruded AZ31B magnesium alloy under different strain ratios, Mater. Sci. Eng. A. 649 (2016) 93–103. https://doi.org/10.1016/j.msea.2015.09.084.

[34] Q. Yu, J. Zhang, Y. Jiang, Fatigue damage development in pure polycrystalline magnesium under cyclic tension – compression loading, Mater. Sci. Eng. A. 528 (2011) 7816–7826. https://doi.org/10.1016/j.msea.2011.06.064.

[35] X.Z. Lin, D.L. Chen, Strain controlled cyclic deformation behavior of an extruded magnesium alloy, Mater. Sci. Eng. A. 496 (2008) 106–113. https://doi.org/10.1016/j.msea.2008.05.016.

[36] J. Segurado, R.A. Lebensohn, J. Llorca, Computational Homogenization of Polycrystals, 1st ed., Elsevier Inc., 2018. https://doi.org/10.1016/bs.aams.2018.07.001.

[37] F. Wang, J. Dong, M. Feng, J. Sun, W. Ding, Y. Jiang, A study of fatigue damage development in extruded Mg–Gd–Y magnesium alloy, Mater. Sci. Eng. A. 589 (2014) 209–216. https://doi.org/10.1016/j.msea.2013.09.089.

[38] F. Yang, S.M. Yin, S.X. Li, Z.F. Zhang, Crack initiation mechanism of extruded AZ31 magnesium alloy in the very high cycle fatigue regime, Mater. Sci. Eng. A. 491 (2008) 131–136. https://doi.org/10.1016/j.msea.2008.02.003.

[39] Z. Wang, S. Wu, G. Kang, H. Li, Z. Wu, Y. Fu, P.J. Withers, In-situ synchrotron X-ray tomography investigation of damage mechanism of an extruded magnesium alloy in uniaxial low-cycle fatigue with ratchetting, Acta Mater. 211 (2021) 116881. https://doi.org/10.1016/j.actamat.2021.116881.

[40] J.D. Bernard, J.B. Jordon, M.F. Horstemeyer, H. El Kadiri, J. Baird, A.A. Luo, Structure – property relations of cyclic damage in a wrought magnesium alloy, Scr. Mater. 63 (2010) 751–756. https://doi.org/10.1016/j.scriptamat.2010.05.048.

[41] M. Zhang, H. Zhang, A. Ma, J. Llorca, Experimental and numerical analysis of cyclic deformation and fatigue behavior of a Mg-RE alloy, Int. J. Plast. 139 (2021) 102885.

https://doi.org/10.1016/j.ijplas.2020.102885.

[42] F. Bachmann, R. Hielscher, H. Schaeben, Texture analysis with MTEX- Free and open source software toolbox, Solid State Phenom. 160 (2010) 63–68. https://doi.org/10.4028/www.scientific.net/SSP.160.63.

[43] Y. Uematsu, T. Kakiuchi, S. Tamano, S. Mizuno, K. Tamada, Fatigue behavior of AZ31 magnesium alloy evaluated using single crystal micro cantilever specimen, Int. J. Fatigue. 93 (2016) 30–37. https://doi.org/10.1016/j.ijfatigue.2016.08.008.

[44] Z. Xing, H. Fan, J. Tang, B. Wang, G. Kang, Molecular dynamics simulation on the cyclic deformation of magnesium single crystals, Comput. Mater. Sci. 186 (2021) 110003. https://doi.org/10.1016/j.commatsci.2020.110003.

[45] D.K. Xu, E.H. Han, Relationship between fatigue crack initiation and activated $\{1\ 0\ \bar{1}\ 2\}$ twins in as-extruded pure magnesium, Scr. Mater. 69 (2013) 702–705. https://doi.org/10.1016/j.scriptamat.2013.08.006.

[46] T.R. Bieler, R. Alizadeh, M. Peña-Ortega, J. Llorca, An analysis of (the lack of) slip transfer between near-cube oriented grains in pure Al, Int. J. Plast. 118 (2019) 269–290. https://doi.org/10.1016/j.ijplas.2019.02.014.

[47] R. Alizadeh, M. Peña-ortega, T.R. Bieler, J. LLorca, Scripta Materialia A criterion for slip transfer at grain boundaries in Al, Scr. Mater. 178 (2020) 408–412. https://doi.org/10.1016/j.scriptamat.2019.12.010.